\documentclass[usenatbib]{mn2e}

\newcommand{\etal}{{et al.}}

\newcommand{\be}{\begin{equation}}
\newcommand{\ee}{\end{equation}}

\newcommand{\kmsMpc}{km~s$^{-1}$ Mpc$^{-1}$}

\newcommand{\mnras}{MNRAS}
\newcommand{\apj}{ApJ}
\newcommand{\aj}{AJ}

\newcommand{\aap}{A\&A}
\newcommand{\apjs}{ApJS}
\newcommand{\araa}{ARA\&A}

\usepackage{epsfig}
\begin{document}

\title[Galaxy Zoo: Bars in Disk Galaxies]{Galaxy Zoo: Bars in Disk Galaxies$^*$ }
\author[K.L. Masters \etal]{Karen L. Masters$^1$, Robert C. Nichol$^1$, Ben Hoyle$^{1,2}$, Chris Lintott$^{3,4}$, \newauthor Steven Bamford$^5$,  Edward M. Edmondson$^1$,  Lucy Fortson$^{4,6}$, William C. Keel$^7$,   \newauthor 
Kevin Schawinski$^8$, Arfon Smith$^3$, Daniel Thomas$^1$ \\
 $^1$Institute for Cosmology and Gravitation, University of Portsmouth, Dennis Sciama Building, Burnaby Road, Portsmouth, PO1 3FX, UK \\
 $^2$Institute for Sciences of the Cosmos (ICCUB), University of Barcelona, Marti i Franques 1, Barcelona, 08024 Spain\\
 $^{3}$Oxford Astrophysics, Department of Physics, University of Oxford, Denys Wilkinson Building, Keble Road, Oxford, OX1 3RH, UK\\
 $^4$Astronomy Department, Adler Planetarium and Astronomy Museum, 1300 Lake Shore Drive, Chicago, IL 60605, USA\\
 $^5$Centre for Astronomy \& Particle Theory, University of Nottingham, University Park, Nottingham, NG7 2RD, UK\\
 $^6$School of Physics and Astronomy, University of Minnesota, Minneapolis, MN 55455, USA\\
 $^7$Department of Physics \& Astronomy, 206 Gallalee Hall, 514 University Blvd., University of Alabama, Tuscaloosa, AL 35487-0234, USA\\
 $^{8}$Einstein Fellow/Yale Center for Astronomy and Astrophysics, Yale University, P.O. Box 208121, New Haven, CT 06520, USA \\
\\ 
\\
 $^*$This publication has been made possible by the participation of more than 200,000 volunteers in the Galaxy Zoo project. \\ Their contributions are individually acknowledged at \texttt{http://www.galaxyzoo.org/Volunteers.aspx}. \\
\\
{\tt E-mail: karen.masters@port.ac.uk}
 }

\date{Accepted for publication in MNRAS, 7th October 2010.}
\pagerange{1--10} \pubyear{2010}

\maketitle

\begin{abstract}

We present first results from Galaxy Zoo 2, the second phase of the highly successful Galaxy Zoo project ({\tt www.galaxyzoo.org}). Using a volume--limited sample of 13665 disk galaxies ($0.01< z < 0.06$ and $M_r<-19.38$), we study the fraction of galaxies with bars as a function of global galaxy properties like colour, luminosity and bulge prominence. Overall, $29.4\pm0.5\%$ of galaxies in our sample have a bar, in excellent agreement with previous visually--classified samples of galaxies (although this overall fraction is lower than measured by automated bar--finding methods). We see a clear increase in the bar fraction with redder $(g-r)$ colours, decreased luminosity and in galaxies with more prominent bulges, to the extent that over half of the red, bulge--dominated, disk galaxies in our sample possess a bar. We see evidence for a colour bi-modality for our sample of disk galaxies, with a ``red sequence" that is both bulge and bar--dominated, and a ``blue cloud"  which has little, or no, evidence for a (classical) bulge or bar. These results are consistent with similar trends for barred galaxies seen recently both locally and at higher redshift, and with early studies using the RC3. We discuss these results in the context of internal (secular) galaxy evolution scenarios and the possible links to the formation of bars and bulges in disk galaxies.    
\end{abstract}


\section{Introduction}
Bars are common in disk galaxies, and are thought to have an important impact on the evolution of galaxies through their ability to transfer angular momentum in both the baryonic and dark matter components of the galaxy \citep{CS81,W85,DS00,B06}. Bars are efficient at driving gas inwards, perhaps sparking central star formation \citep[e.g.][]{H86,K95,J05,S05}, and thus help to grow a central bulge \citep[e.g. for a review][]{KK04}. Bars may also feed a central black hole (as per \citealt{S89,S90} and see \citealt{J06} for a recent review), but so far no correlation has been found between AGN activity and bar fraction in galaxies forming stars (e.g. \citealt{H97} and as recently discussed by \citealt {H09}). 

Early visual inspection of spiral galaxies in catalogues like the RC3 \citep{RC3} gave an optical bar fraction of $f_{\rm bar} \sim 0.25-0.3$, rising to 60\% if weaker bars or oval distortions were included \citep[as discussed in e.g.][]{SW93,M95,KSP00,S08}

More recent work on the bar fraction of disk galaxies has relied on automated methods of detecting bars in galaxies including elliptical isophote fitting or the Fourier decomposition of CCD images. Such automated studies find optical bar fractions of $\sim$50\% for nearby disk galaxies \citep{Bar08,Ag09}, which is consistent with near infra-red (NIR) studies that find a majority (at least 60\%) of disk galaxies appear to have a bar \citep[e.g.][]{MR97,MJ07}. These differences are probably due to a combination of selection effects, wavelength-dependences, differences in the strength of the bar, and small samples sizes. While studies of bars in galaxies have been collectively moving towards automated classifications (and away from possibly subjective visual classifications) in recent years, concerns have arisen about the reliability of a completely automated approach for detecting bars which struggle to distinguish spiral arms and bars in some cases \citep[see for example the discussion of problems with the Fourier method in][]{Ag09}.

For such reasons, especially to provide a larger sample of visually selected barred galaxies, we started a new phase of the successful Galaxy Zoo project\footnote{{\tt www.galaxyzoo.org}} by asking the public to provide more detailed visual classifications of galaxies seen in the Sloan Digital Sky Survey (SDSS). This new project is known as ``Galaxy Zoo 2" (GZ2) throughout this paper. The project and data set will be fully described in a future paper (Lintott et al. in prep.).

In this article, we present the first results on the bar properties of 13665 visually classified GZ2 disk galaxies. This sample is nearly an order of magnitude larger than previous studies using SDSS data \citep{Bar08,Ag09} which facilitates a detailed statistical study of the fraction of barred disk galaxies as a function of other galaxy properties like  global optical colour, luminosity, and estimates of the bulge size, or prominence. Where appropriate, we assume a standard cosmological model of $\Omega_{m}$ = 0.3, $\Omega_{\Lambda}$ = 0.7 and $H_{0} = 70$ \kmsMpc and all photometric quantities are taken from SDSS.

\section{Identification of Bars and Sample Selection}

\begin{figure*}
\includegraphics[height=120mm,angle=-90]{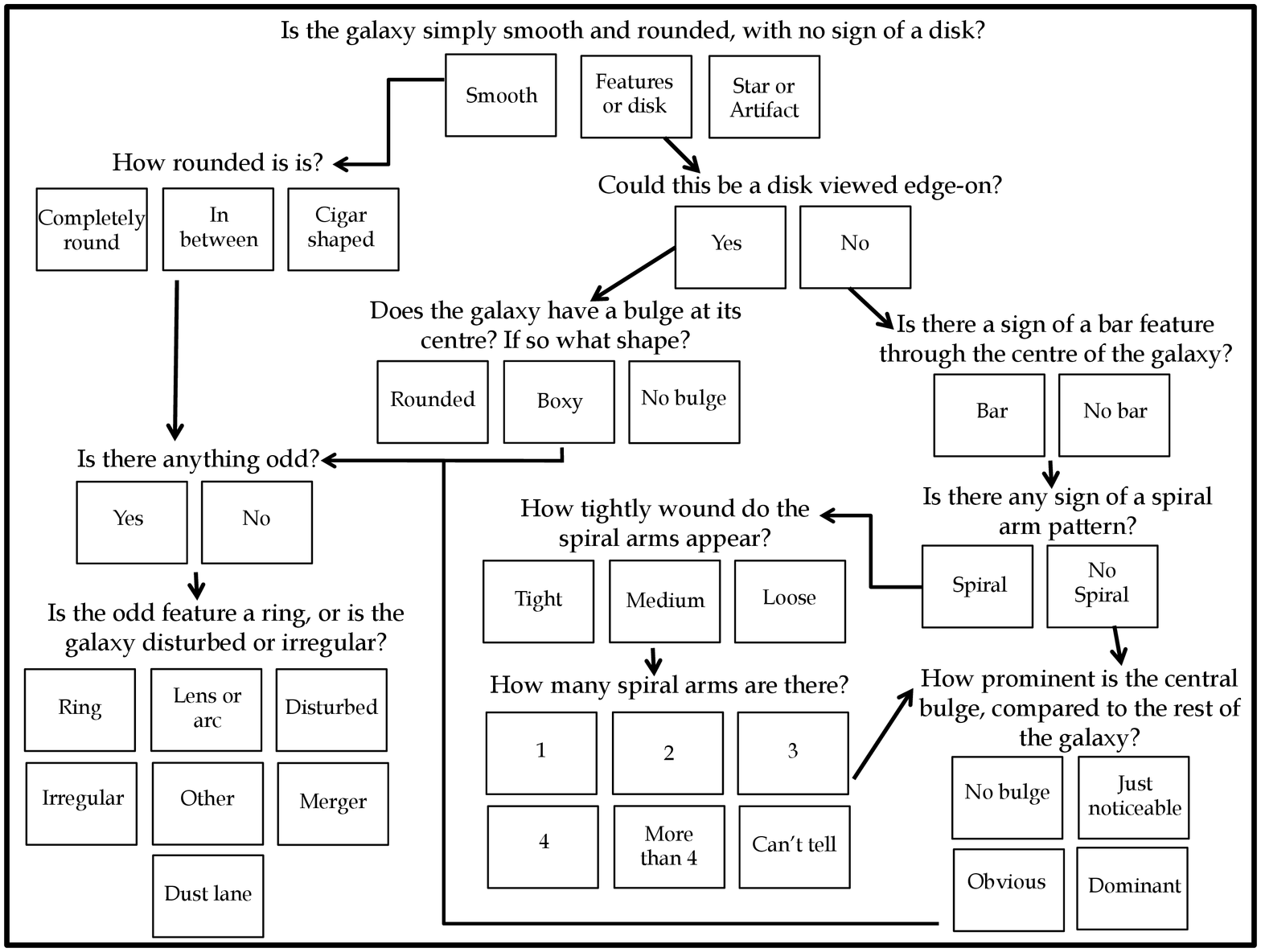}
\caption{We present a schematic diagram of the decision tree for GZ2 classifications. We provide the questions asked of the user for each SDSS galaxy image (starting with the top question first).  For each question, we provide the possible answers they are allowed. Depending on their answers, the user can navigate down different branches of the tree. \label{GZ2}}
\end{figure*}

The first phase of Galaxy Zoo\footnote{\tt http://zoo1.galaxyzoo.org} (known as GZ1 hereafter) has now finished collecting data, is described in detail in \citet{L08} and the data have been made public \citep{L10}.  In the second phase (GZ2), users were asked to provide more detailed classification of galaxies  than in the original GZ1 (see Figure \ref{GZ2}). Specifically, after identifying a galaxy as possessing {\it ``features or a disk"} (see top question in Figure \ref{GZ2}), the users were then asked different questions (depending on their prior answers) as they navigate to the bottom of the decision tree presented in Figure \ref{GZ2}. For example, if the user identified a disk from the first question in the tree, they are then asked if the galaxy could be an edge-on disk, and if the answer to that question is ``no'', then they are further asked to identify if the galaxy has a bar or not. This identification process is based on the SDSS $gri$ composite images as it was in GZ1. 

At the time of submission of this paper we were still collecting data for the GZ2 project, and these first results are based upon data collected up to July 2009. Only galaxies for which at least ten answers to the bar question (the question beginning {\it ``Is there a sign of a bar feature...."} in Figure \ref{GZ2}) have been included in the sample, which gives a total superset of 66,835 disk galaxies in which the median number of ``bar" classifications is 20. In GZ2 galaxies were presented to the user randomly for classification. However only galaxies in which a user identified as having {\it ``features or a disk"} (see Figure \ref{GZ2}) progressed to the question about bars, so galaxies possessing 10 answers to this question are likely to be biased toward later types but can include early types if any identifiable features (e.g. bars) are present. 

For the rest of this paper, we classify a galaxy as being ``barred" if the number of users identifying them as having a bar is equal to, or larger than, the number identifying them as not having a bar, i.e., a majority of users voted they saw a bar. A fraction of the sample has also been visually inspected by us and this choice seems to make sense (see Figure 2). However, as has been discussed extensively for GZ1 data \citep[e.g.][]{L08,B09}, there are many ways to go from ``clicks to classifications". This simple choice means that no galaxy is left unclassified, and gives equal weight to all users. Alternative threshold classifications were explored, and were found to make no qualitative difference to the results. Future studies using GZ2 data will no doubt explore this issue further. 

\begin{figure*}
\includegraphics[height=150mm,angle=-90]{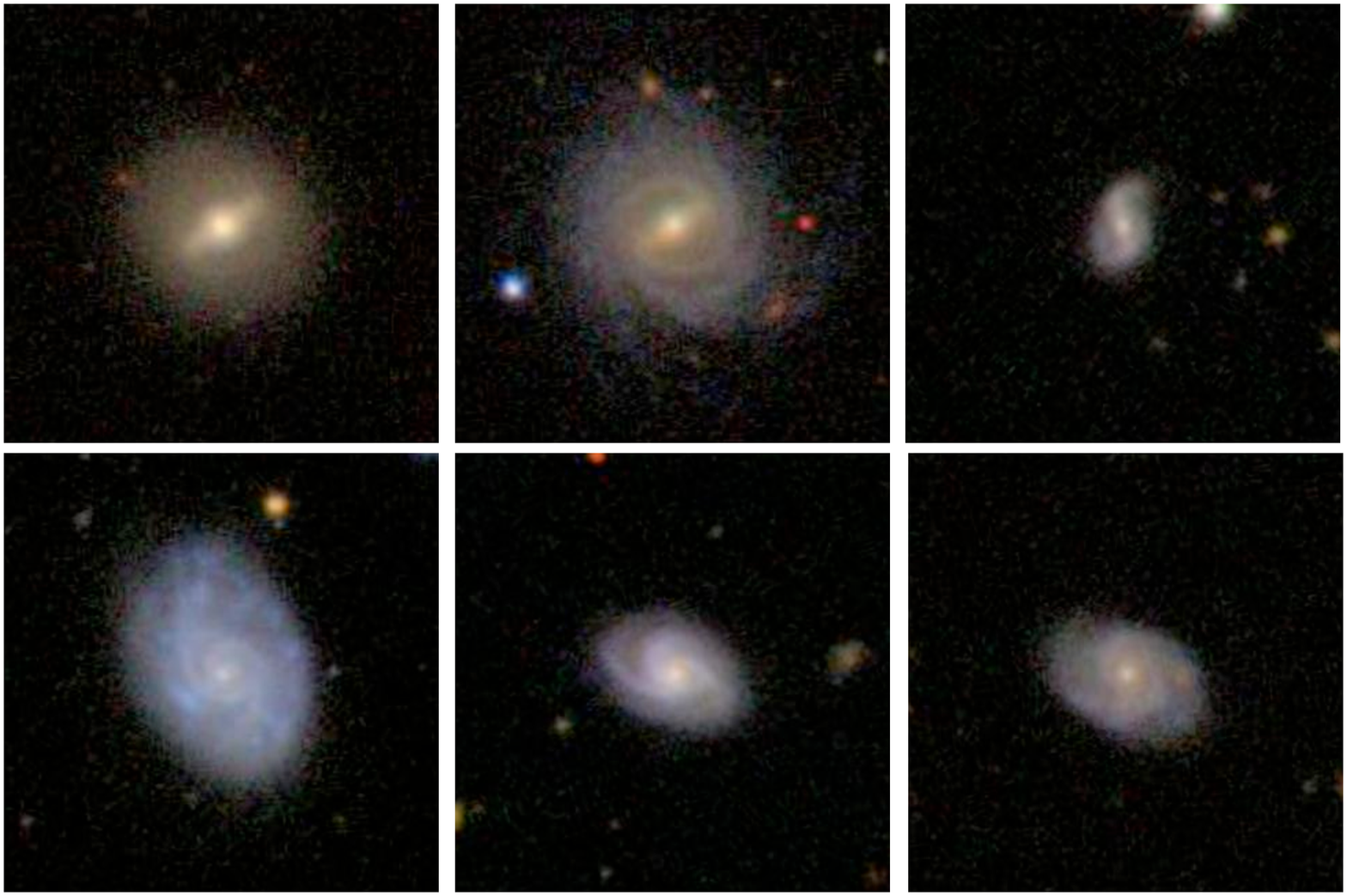}
\caption{(Top row) Examples of GZ2 classified barred disk galaxies. (Bottom row) Examples of GZ2 classified disk galaxies with no bar. The galaxies on the left are at $z\simeq0.02$, the galaxies in the middle at $z\simeq0.04$ and the galaxies on the right are at $z\simeq0.06$, thus spanning the full redshift range of the volume--limited sample used herein (see Section 2).  The images are taken from the SDSS ($gri$ composite) and are one arcminute squared in size. (These images differ to those presented to users for classification, which are scaled using the Petrosian radius of the galaxy.) 
\label{examples}}
\end{figure*}

We select from the superset of 66,835 disk galaxies a volume--limited subsample of GZ2 disk galaxies with $0.01<z<0.06$ and $M_r<-19.38$, where $M_r$ is the SDSS Petrosian r-band magnitude k-corrected to z=0 and with the standard Galactic dust extinction correction (Schlegel, Finkbeiner \& Davis 1998). A correction for dust extinction internal to the galaxy, as described in \citet{GZdust}, is also applied (this corrects for the inclination dependence and is zero for face-on disks). Furthermore, we limit our sample to $\log (a/b) < 0.3$ ($i\sim60^\circ$)  as identifying bars in highly inclined disk galaxies is challenging. This is a comparable cut in inclination to previous studies of bars in spiral galaxies \citep[e.g.][]{S08,Bar08,Ag09}. These constraints provide a final sample of 13665 disk galaxies used throughout this paper (with median number of 22 answers to the GZ2 ``bar" question). Examples of barred and non--barred galaxies in our sample are shown in Figure \ref{examples}, over the range of redshifts included in this study. We have cross-matched this sample with the GZ1 (SDSS DR6) sample discussed in \citet{B09}. The distribution of GZ1 spiral likelihood, $p_{\rm sp}$ for the disk galaxies in this sample peaks at $p_{\rm sp}=1$ with a median value of $p_{\rm sp}=0.9$, and a long low tail to small spiral likelihoods. Overall 66\% of the sample would be classifed by GZ1 as spiral ($p_{\rm sp} > 0.8$) with less than 1\% GZ1 early types ($p_{\rm el}>0.8$). The remaining fraction have uncertain classifications in GZ1 (using the ``clean" sample criteria). This cross-match indicates that what we will call ``disk galaxies" in GZ2 should be interpreted as being mostly classic spiral galaxies, but also consisting of earlier type galaxies in which a ``disk or features"  (see the top question of Figure 1) was discernable to GZ2 users. 

The stellar mass range of this volume-limited GZ2 sample of 13665 disk galaxies is approximately $10^9 < M < 10^{11} M_\odot$  (using the stellar mass estimates from Baldry et al. 2006) with a colour and luminosity dependence such that the dimmest, bluest objects in our sample have stellar masses of $10^9 < M < 10^{10} M_\odot$, while the reddest, most luminous, galaxies have higher stellar masses (red galaxies in our sample are only complete to stellar masses of $10^{10} M_\odot$). 

 To address concerns that the requirement of $n>10$ answers to the GZ2 bar question might bias the sample we compare the luminosity, colour, axial ratio and redshift distributions compared to GZ1 selected spirals ($p_{\rm sp}>0.8$) in the same volume limit and find no significant differences in these distributions excepting a slightly larger tail to more luminous, redder and more distant galaxies in the sample studied here - easily explained by the addition of small numbers of early types. It might initially appear worrisome that early types may be present in our sample if they possess a bar (since then they make it past the``features or a disk" question) but not if they are unbarred. However the contamination appears small enough that we find no differences in the results presented below if the sample is further restricted to include only GZ1 ``clean" spirals or if we remove the small fraction of GZ1 ``clean" early -types. Therefore we leave the selection as is.

\section{Results}

First, we compute the overall mean bar fraction of our sample which is $29.4\pm0.5$\% (we find 4020 barred disk galaxies in our sample of 13665 GZ2 disk galaxies). This value is in excellent agreement with the fraction of 25-30\% of galaxies having strong bars found by visual inspection of classic optical galaxy samples \citep[e.g. the RC3 and UGC][]{RC3,UGC} and also with Marinova et al. (2009) who found a bar fraction of $\sim$30\% for disk galaxies in a dense cluster at $z\sim0.165$ (in the STAGES survey). However, it is lower than recent bar fractions quoted for nearby disk galaxies using automated ellipse fitting techniques to find bars, e.g., \citet{Bar08} find a bar fraction of $50\pm2$\%, while \citet{Ag09} find 45\%. This difference could depend on the strength of bars included in these analyses as it has been known for sometime that the bar fraction in the RC3 catalogue increases to $\sim 60\%$ if weak or ovally distorted systems are included \citep[e.g.][]{SW93,KSP00}. We return to this issue in Section 3.1 and in future work, but it appears our GZ2 bars may be consistent with the classic optically--identified strong bars (ie. SB types). 
 
We see no trend of bar fraction with redshift in our sample, and only a mild trend with inclinations (for $i<60^\circ$) such that bars are slightly less likely to be identified as the galaxies become more inclined. As the inclinations are random with respect to other galaxy properties, we do not expect this trend to have a significant effect on our results.  We argue here that the absence of a trend of bar fraction with redshift may also be taken as suggestive that our bar classification using a simple $N_{\rm bar} > N_{\rm no bar}$ from GZ2 clicks picks out only the strongest and largest bars, since smaller bars would not be resolved over the full redshift range considered. We reiterate that all trends of bar fraction we discuss should most likely be considered trends in the fraction of strong or obvious bars. Future work comparing GZ2 bar classifications with other classification methods will test this. 
 
\subsection{Bar Fraction with Colour}

\begin{figure}
\includegraphics[width=84mm]{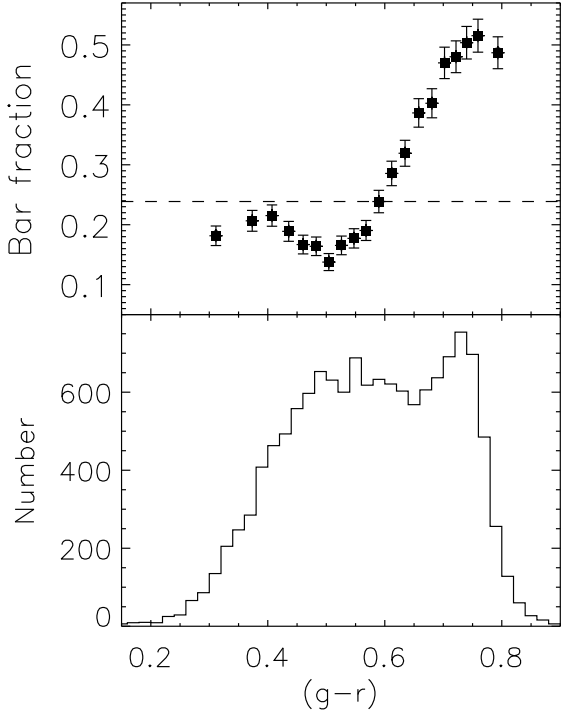}
\caption{(Top panel) The bar fraction as a function of global galaxy $(g-r)$ colour. The dashed line shows the median bar fraction for the entire volume--limited sample of GZ2 disk galaxies. Poisson error bars are shown. (Lower panel) The distribution of $(g-r)$ colours for GZ2 disk galaxies used in this study. \label{barfraction_colour}}
\end{figure}

In Figure  \ref{barfraction_colour}, we present the bar fraction of our GZ2 volume--limited sample as a function of the $(g-r)$ global, k-corrected colour of the galaxy\footnote{Standard Galactic extinction corrections are also applied, but no correction for internal extinction -- this last corrections would be at most 0.04 mag because of the inclination limit $\log (a/b) <0.3$}. We find a significant trend for redder GZ2 disk galaxies to have larger bar fractions, e.g.,  over half of the reddest galaxies in our sample possess a bar. This is consistent with the recent results of  \citet{M10} who found that passive red spiral galaxies had a high fraction of bars; these passive spirals would be in the extreme red population in our GZ2 sample. Interestingly, the trend seen in Figure 3 does not appear to be monotonic, given the Poisson error bars, and we see a slight increase in the bar fraction for the bluest objects (compared to intermediate colours of $(g-r)\sim 0.5$). 

Furthermore, one could argue that the trend seen in Figure 3 is consistent with a difference in bar fraction which is correlated with the well-established colour bi-modality relationship of galaxies (illustrated in the lower panel of Figure \ref{barfraction_colour}).  Disks in the ``blue cloud" (e.g., with $(g-r) < 0.6$) have a constant bar fraction (with colour) of $\simeq 20\%$, while disk galaxies in the ``red sequence" (e.g., $(g-r) > 0.7$) have a clear increase in bar fraction with colour.  The split between ``normal" and ``red" passive spiral galaxies discussed by \citet{M10} falls approximately at the dip of the colour distribution of our GZ2 disk galaxies (the colour cut applied by \citet{M10} was $(g-r) = 0.63 - 0.02(M_r + 20)$ while the median magnitude of our GZ2 sample used here is $M_r = -20.9$, but note that \citet{M10} also removed spiral galaxies with any sign of a bulge, something which has not been done here). 

We should consider if the trend of bar fraction with colour could be an artifact of the visual identification of bars in the SDSS composite $gri$ images. It does seem feasible that the bluer GZ2 disks might have stellar bars hidden under the on--going star formation which would dominate the SDSS $g$-band, while in the redder disks, the light would be dominated by the $i$-band, and thus lead to an increased bar fraction. It has been argued in the literature that bar fraction increases when moving from the optical into the NIR \citep[e.g.][]{E00,MJ07}, with more bars being revealed in the NIR \citep[e.g.][]{K96,M97}. However more recently it has been quite clearly shown that overall the bar fraction does not change between optical (B-band) and the NIR \citep[e.g.][]{E02,MD07}. In particular we highlight the findings of \citet{S08} who show for a sample of 139 local SDSS galaxies, that bar fraction (both from visual inspection and ellipse fitting) is constant over the SDSS $griz$ passbands, and only drops significantly in the $u$-band. Furthermore, we highlight the fact that the bar fraction for our bluest GZ2 disk galaxies ($(g-r)<0.5$) is greater than at intermediate colours (as discussed above) and is consistent with a constant bar fraction for all colours bluer than $(g-r)<0.6$. These results argue against a significant colour bias in identifying bars in our composite $gri$ images but further studies may be necessary to discover any subtle biases with colour. 

\subsubsection{Bar Fraction, Colour and Luminosity}

\begin{figure}
\includegraphics[width=84mm]{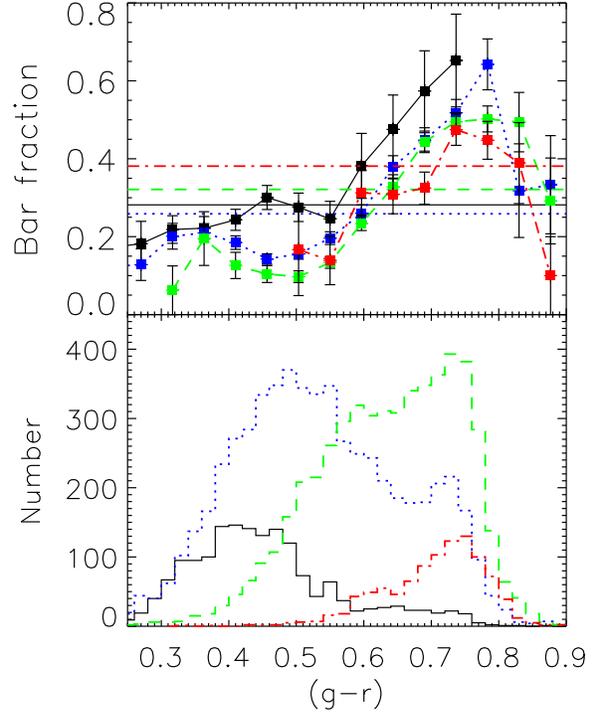}
\caption{Similar to Figure \ref{barfraction_colour}, but with the sample split into four bins of absolute magnitude: $M_r > -20$ (black solid), $-20< M_r < -21$ (blue dotted), $-21 < M_r < -22$ (green dashed) and $M_r < -22$ (red dot-dashed)\label{barfraction_colour_magnitude}. Bins are only plotted if they have at least 10 galaxies.}

\end{figure}

The trend observed in Figure \ref{barfraction_colour}, and in particular the upturn in bar fraction at the bluest colours, suggests that colour may not be the only variable of importance for the bar fraction.  We therefore split the sample into four subsamples of absolute magnitude and show the results in Figure \ref{barfraction_colour_magnitude} (we avoid using stellar mass estimates which introduce completeness effects dependent on colour). This figure shows that at a fixed $(g-r)$ colour, there is a residual trend of bar fraction with luminosity such that bars are more common in lower luminosity disk galaxies (since this is at fixed colour this corresponds to lower stellar mass). However, the trend with luminosity is still sub--dominant compared to the correlation of bar fraction with colour. 

Figure \ref{barfraction_colour_magnitude}  highlights a  maximum bar fraction of $\sim 70\%$ (for our optical GZ2 sample) for low luminosity disk galaxies with colours in the range of $0.7<(g-r)<0.8$, and a drop in bar fraction for the most massive, red disks in this sample (if these are identified with S0s this drop in bar fraction has been observed before, e.g. \citealt{Lauri09}). Interestingly the overall trend is for an increase in bar fraction with galaxy luminosity, but this is obviously driven by the trend of bar fraction with colour and the fact that more luminous disk galaxies tend to be redder.

\subsection{Bar Fraction and Bulge Prominence}

As part of GZ2, users were also asked to identify the size, or prominence, of the bulges in disk galaxies (excluding edge-on disks), and were given the four options of ``no bulge", ``just noticeable", ``obvious" and ``dominant" (see Figure 1 for details). Similar to our treatment of the bar question, we uniquely place all of our GZ2 disk galaxies into one of these four categories based on majority voting. We find that most galaxies are placed into the middle two categories, with only a small number of disks having dominant or no bulge classification. In \citet{GZdust}, the use of the SDSS parameter {\tt fracdeV}\footnote{The fraction of the best fit light profile which comes from a de Vaucouleurs fit as opposed to an exponential fit} was used as a proxy for bulge size in GZ1 spirals. In bright spirals, {\tt fracdeV} is dominated by the inner light profile and should be increased in the presence of a large bulge component. 

\begin{figure}
\includegraphics[width=84mm]{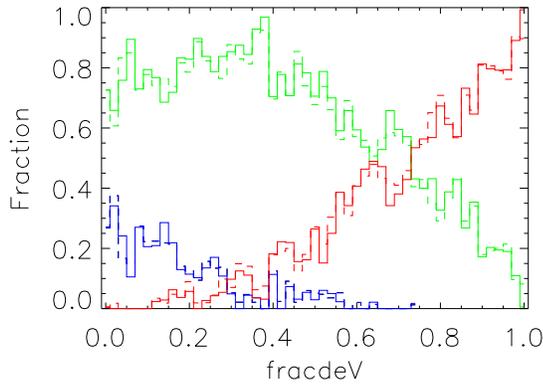}
\caption{The distribution of values of {\tt fracdeV} for GZ2 disk galaxies classified by their visual bulge size into ``no bulge" (blue), ``just noticeable" (green) and ``obvious" (orange). We have not plotted ``dominant" bulges as they are few in this category and a vast majority have {\tt fracdeV}$=1$. Histograms for barred galaxies are shown by the dashed lines, unbarred by the solid lines.\label{pbulge_fdeV}}
\end{figure}

In Figure \ref{pbulge_fdeV}, we show the {\tt fracdeV} distribution of GZ2 disk galaxies separated into the four GZ2 bulge categories. We find that most GZ2 disks with low values of {\tt fracdeV} are categorised as having ``just noticeable" bulges by the GZ2 users (the green line in Figure \ref{pbulge_fdeV}), while most GZ2 disk galaxies with large values of {\tt fracdeV} are categorised as having an ``obvious" bulge (orange line). While the number of galaxies in the ``no bulge" (blue line) and ``dominant" bulge categories are small, there is still a clear trend with {\tt fracdeV} in the expected direction.  These results are reassuring as they demonstrate that both {\tt fracdeV} and the GZ2 bulge classifications are meaningful and do provide a measure of the bulge prominence in disk galaxies. We further split the sample into barred and non-barred disks using the GZ2 classifications to check that the presence of a bar does not have a significant effect on {\tt fracdeV}. The resulting histograms (dashed and solid lines in Figure \ref{pbulge_fdeV}) are identical thus proving there is no impact on {\tt fracdeV} from the presence of a bar. 

\begin{figure}
\includegraphics[width=84mm]{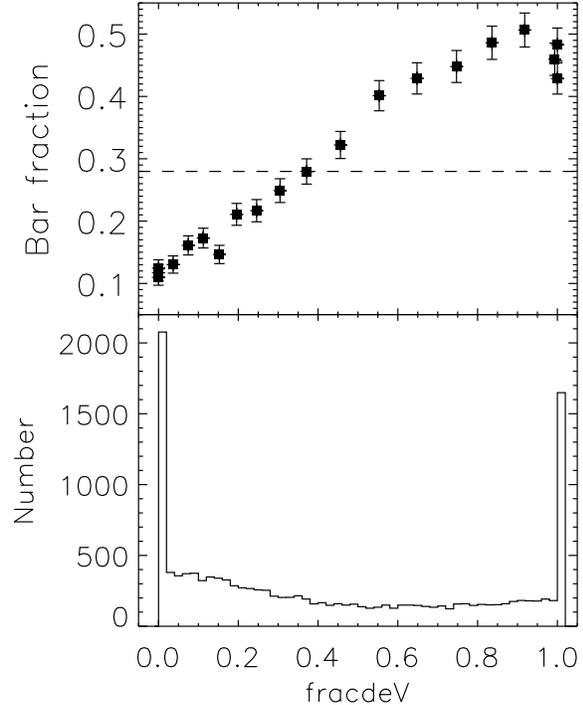}
\caption{(Top panel) The bar fraction as a function of {\tt fracdeV}. The dashed line shows the median bar fraction for the entire volume limited sample of GZ2 disks. Poisson error bars are shown. (Lower panel) The distribution of {\tt fracdeV} values of GZ2 disks used in this study. Most galaxies have either {\tt fracdeV} = 0 or {\tt fracdeV} = 1. \label{barfraction_fracdeV}}
\end{figure}

In Figure  \ref{barfraction_fracdeV}, we show the fraction of GZ2 disk galaxies in our sample as a function of {\tt fracdeV}. We have chosen to use this SDSS measured quantity, instead of the GZ2 bulge classification, to follow the work of  \citet{GZdust} and because it is a continuous variable. Figure  \ref{barfraction_fracdeV} shows a clear monotonic increase of bar fraction with {\tt fracdeV}, i.e,. half of the bulge--dominated disk galaxies in our sample have bars. 
 
\begin{figure}
\includegraphics[width=84mm]{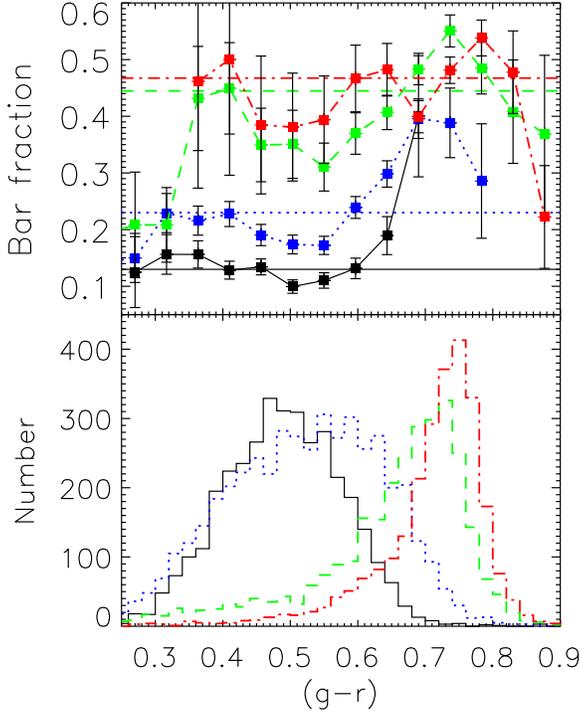}
\caption{Similar to Figure \ref{barfraction_colour} but split into four bins of {\tt fracdeV}: {\tt fracdeV} $<0.1$ (no bulge; black solid lines), $0.1< ${\tt fracdeV} $<0.5$ (small bulge; blue dotted lines), $0.5< ${\tt fracdeV} $<0.9$ (large bulge; green dashed lines) and {\tt fracdeV} $>0.9$ (dominant bulge; red dot-dashed lines).
\label{barfraction_colour_bulge}. Bins are only plotted if they have at least 10 galaxies.}
\end{figure}

As is well known, and recently shown for GZ1 spirals by \citealt{GZdust}, early--type spiral galaxies with large bulges tend to be redder than late--type spirals, so the observed trend in Figure \ref{barfraction_fracdeV} could be due to a correlation between colour and bulge size. To explore this, we split the trend of bar fraction with colour (seen in Figure 3) into four broad disk galaxy types of
\begin{itemize}
\item{no bulge present with {\tt fracdeV} $<0.1$,}
\item{small bulge with $0.1< ${\tt fracdeV} $<0.5$,}
\item{large bulge with $0.5< ${\tt fracdeV} $<0.9$, and}
\item{dominant bulge of {\tt fracdeV} $>0.9$.}
\end{itemize}
The results of this division are shown in Figure \ref{barfraction_colour_bulge}, which shows that late-type disk galaxies (with low {\tt fracdeV}) have a low bar fraction (except the very reddest as also seen in Masters et al. 2010b) while early-type disks have a high bar fraction. This shows that the bar fraction correlation with colour is primarily driven by bulge prominence.

At this point, it is important to recognise the different types of bulges that are observed to be present in disk galaxies \citep{A05a}. In particular, there is a distinction in the literature between {\it classical} bulges and {\it pseudo-}bulges. The former appear to resemble a classic elliptical galaxy which just happens to be within a disk (and are often assumed to be formed by merger events) while the latter are disk-like bulges more consistent with being formed by the re-distribution of material within the disk. These  {\it pseudo-}bulges as they are called have almost exponential profiles \citep{KK04}, while classical bulges more closely follow a de Vaucouleur profile, however (as is seen in elliptical galaxies themselves) a range of profile shapes is observed in bulges which varies with the luminosity of the bulge \citep[][and references therein]{GW08}, and since dEs can also form with almost exponential profiles, pseudo-bulges may not necessarilly be formed by secular evolution \citep{GW08}.

Our use of {\tt fracdeV} as a proxy for bulge size is likely to be most effective in selecting classical bulges (although any central excess over the disc's exponential profile would favour a larger {\tt fracdeV}). In the bottom panel of Figure 7, we see that most of our GZ2 disk galaxies with large {\tt fracdeV} values have red $(g-r)$ colours (also as discussed in \citealt{GZdust}). It is well known that pseudo-bulges are more common in bluer later type spirals while redder early type spirals are more likely to host classical bulges \citep[e.g.][]{KK04,DF07} and although it's over-simplistic to say that all late type spirals have pseudo-bulges and all early types have classical bulges \citep{GW08}, there is clearly a trend.

Therefore, we tentatively interpret our observed trends of increased bar fraction with {\tt fracdeV} and $(g-r)$ colour as hinting that early type red disk galaxies (preferentially with a classical bulge) have a higher fraction of bars than later type blue disk galaxies (preferentially with pseudo-bulges or no bulge). 
    
\section{Discussion}

\subsection{Comparison with Other Work}

 The literature on bar fractions is extensive going back to seminal early work on optical galaxy catalogues. Many studies have attempted to separate the bar fraction into galaxies of different morphological types \citep[e.g.][]{O96,KSP00,E04} and a picture is emerging which suggests that the bar fraction found in a given sample might depend quite sensitively on the morphological and stellar mass make up of the sample being considered \citep[as recently discussed by][]{G10,NA10}. One must of course be careful in comparing {\bf any} galaxy properties across studies using samples at different redshifts and with different morphological and luminosity/mass distributions, but a review of results from different studies is still useful to put our results into context. 
  
In this work, we observe a significant increase in the bar fraction of disk galaxies as the galaxies become redder and have more prominent bulges. We also observe a small increase in the very bluest galaxies in our sample. This suggests that the bar fraction is highest in early type disks, has a minimum somewhere in the blue cloud of spiral disks, and increases slightly towards the latest type spirals. 

In fact there has been a suggestion for some time that bar fraction does not vary monotonically with galaxy type. Both \citet{O96} and \citet{E04}, use visual classifications of RC3 galaxies observe that the (strong) bar fraction (i.e.  spiral types SB) decreases from around 60\% in S0/a to 30\% in Sc, after which it increases again towards very late type disks. Interestingly both studies also show that the weak (or mixed type, SAB) bar fraction is much flatter with Hubble type and if anything shows the opposite trend.  \citet{KSP00} used the same RC3 data to argue that bar fraction remains relatively flat across all types of disk galaxies, however the trend they show in their Figure 1 (for strong bars at least) appears consistent with that used by both \citet{O96} and \citet{E04} to argue for variation. 

Almost concurrently with this work, \citet{G10} in a multiwavelength study of galaxies in the Virgo cluster, show that the bar fraction depends sensitively on the morphological composition of a sample. In qualitative agreement with our results, they find in a sample of $\sim 300$ disk galaxies that early type spirals have a higher bar fraction (45-50\%) than late-type spirals (22-36\%) and suggest the difference could be explained by the higher baryon fraction of earlier type spirals. 

Also in very recent work, \citet{NA10} discuss the bar fraction in 14,043 visually classified galaxies (all classified by PN, \citealt{NA10a}) and like us find the bar fraction to depend on morphology with a minimum near the division between the blue and red sequences. They suggest that to reconcile the apparently conflicting results on the bar fraction (and in particular the evolution of the bar fraction with redshift) found in the literature one need only consider the different stellar mass ranges of the samples in question.

Other recent work on the bar fraction at low redshifts ($z < 0.1$ or so) also support this broad picture of bar fraction having a minimum at around Sc types and rising towards both earlier and later spiral/disk galaxy types. For example, \citet{Lauri09} study the bar fraction in 127 early type spirals and find it increases from Sas to S0/a (but then drops significantly in S0s - however distinguishing unbarred S0s from elliptical galaxies is notoriously hard and may bias the S0 bar fraction low). At first glance our results appear in conflict with those of \citet{Bar08, Ag09, We09} all of who argue that bar fraction increases as bulge prominence decreases, however we suggest that sample selection may again be the culprit, with these studies actually picking out significantly later spiral types than are found in our sample, and therefore seeing only the bluest end of the trend we show. 

For example \citet{Bar08} and \citet{Ag09}, who both find larger bar fractions in later/bluer disk galaxies (using a local samples of $\sim2000$ ``disk" or spiral galaxies from the SDSS)  both use automated techniques to identify a disk/spiral sample of galaxies based on concentration and velocity dispersion \citep{Ag09} or colour \citep[][this study also considered Sersic fits, but the final results were for a colour--selected sample of ``spiral" galaxies]{Bar08}. If we restrict our analysis to the range of colours explored by \citet{Bar08}, then much of the trend we see in Figure 3 is missed and we would have actually witnessed a mild decrease in bar fraction towards the redder spirals, fully consistent with their findings. In more detail, we mimic the \citet{Bar08} selection by using their U-V colour cut (from \citet{B04}) with colour transformations from \citet{S02}, plus $i<60^\circ$, and $0.01<z<0.03$, $M_g<-18.5$. For this matched GZ2 sample,  we find a flat bar fraction of 25\%, with no obvious trend with $(g-r)$ colour, except for a slight upturn for the reddest objects (at $(g-r) \sim 0.6$).

Comparing our results to those at higher redshifts must be done with caution (considering the different stellar mass ranges considered - we remind the reader that our sample consists of disk galaxies with stellar masses between $10^9$ and $10^{10}  M_\odot$). However it is interesting that we observe a similar trend of bar fraction to that shown in the COSMOS sample ($z\sim$ 0.2--0.8) by both \citet[][in 2157 spiral galaxies]{S08} and \citet[in 3187 disk galaxies]{C10}. Both these studies find more bars in redder disk galaxies at intermediate stellar masses of $10^{10.5}<M<10^{11}M_\odot$, which is similar to the high end of the mass range of our sample. We note that \citet{C10} finds this trend changes at masses$>10^{11}M_\odot$, but there are almost no such galaxies in our volume--limited sample. 

   While we find a similar overall bar fraction to the STAGES study of
barred galaxies in a dense cluster at $z\sim0.2$ (Marinova et al.
2009) our findings on the trends of bar fraction with other properties
differ substantially from that study. They observed that bar fraction
(in $\sim 800$ galaxies found from ellipse fitting to B-band images) rises in brighter galaxies and
those which have no significant bulge component and that bar fraction
had no dependence on disk galaxy colour. While we do see an increase
in bar fraction for brighter spirals (from $26\pm1$\% for those with
$M_r>-20$, to $37\pm2$\% for those with $M_r<-22$), we argue that this
is driven by the strong colour dependence of the bar fraction and at a
fixed colour we see little dependence of bar fraction on luminosity
(Figure 4) - the trend we find is also much smaller than seen by
Marinova et al. (2009). 

Similarly, we agree with the overall bar fraction of 25\% found for 945 galaxies by \citet{Bar09} across both field and clusters environments at $z\sim$0.4--0.8 (observed in rest frame B--V). However we again find opposite trends of bar fraction with bulge-prominence (they find more bars in bluer disk-dominated galaxies). 

The source of these discrepancies is unclear. Disk galaxies were identified
visually in both studies so should be similar to our GZ
disks. We do use quite different bar finding techniques (visual versus
ellipse fitting), so perhaps this indicates a difference in the trends
for strong (visual) and weaker bars (as also hinted at by \citealt{O96} and \citealt{E04}). More interestingly it could be
pointing to a difference between disk galaxies in high density regions 
and those elsewhere (as also discussed by \citet{G10} who find similar results to us in Virgo cluster galaxies). \citet{Bar09} explored differences between bar fractions in the field and clusters finding hints that high density regions are favourable locations for bars, however this could only be done for a subset of the sample ($N=241$), making the results of limited statistical significance. These possibilities will be explored in future work exploiting the huge number of bar classifications available to us in GZ2.

There does remain a difference in our overall conclusions with \citet{Bar08} and \citet{Ag09} and, in particular, the total bar fraction we find is lower than either of these studies. It has been argued that bars in early-type spiral galaxies tend to be longer (and thus stronger) than those seen in late--type spirals \citep{A03}, so it may be that the remaining differences are due to our sensitivity to these longer, stronger bars, while \citet{Bar08} and \citet{Ag09} may detect weaker bars using their ellipse--fitting techniques.  The cross-over between the sample used here and that in \citet{Bar08} is small (as we have argued above it is this mismatch which explains the different trends we see in bar fraction with colour and bulge prominence), but in $\sim$400 galaxies found in both samples, we find GZ2 bars in $\sim$ 25\% of the objects while the number of barred objects rises to roughly twice that in the \citet{Bar08} classifications. The two bar classification methods agree $\sim$75\% of the time -- most of the difference is from \citet{Bar08} identified bars not being found by GZ2.    Further studies directly comparing the bars identified from ellipse--fitting methods and the GZ2 identifications will be needed to understand the main reason for this difference. Such a comparison is in progress (Masters et al. in prep.). However, it is interesting that the bar fraction difference between the two techniques is rather similar to the split between strongly barred (SB), weakly barred (SAB) and non-barred (SA) galaxies in both the RC3 \citep[e.g.][]{SW93,E00} and the de Vaucouleurs Atlas of Galaxies (Buta, Cordwin \& Odewahn 2007). This seems to further indicate that (as we suggested above) GZ2 bars should be identified with strong bars (SB types) only and so the trends we observe should most likely be considered trends in the fraction of strong bars.

\subsection{The Impact of Bars on Disk Galaxies}

Given the trends we have observed, we now focus on the interpretations we can make for the effect of bars on the secular and dynamical evolution of disk galaxies. We observe a significant increase in bar fraction as disk galaxies become redder and have larger (classical) bulges; over half of red, bulge--dominated disk galaxies have a bar. 

Our observations suggest an important link between the presence of a bulge (perhaps preferentially a classical bulge with a de Vaucouleur profile) and the existence of a bar instability. Bar instabilities are often invoked as a way to form pseudo-bulges (with an exponential profile) by moving material around in the disk of a spiral galaxy (see \citet{KK04} for a comprehensive review of this subject), but classical bulges are usually thought to have formed during a fast, dissipative process (most likely related to galaxy mergers), which would have likely disrupted any bar. However using a sample of 143 local galaxies with bulge-disk-bar decompositions, \citet{We09} argue that most bright spirals have (almost) pseudo-bulges and comparing with cosmological simulations argue that most of these must have been made by a combination of minor mergers and secular evolution. Perhaps it is this link we are seeing in the Galaxy Zoo sample. 

In \citet{G10} the observation that the barred fraction is higher in early type disk galaxies in their Virgo cluster sample is tentatively explained by a combination of the higher baryon fraction in early type galaxies (which makes their discs heavier and therefore more susceptible to bar instabilities) and environmental effects which might destroy a late-type spiral but would leave an early type spiral with a bar. Further studies of the environmental dependence of the bar fraction are planned with GZ2 data (Skibba et al. in prep.) and will be used to test this scenario. 

We finish by returning to the suggestion that we see two populations of disk galaxies. Both Figure 3 and 7 suggest a split between disk galaxies on the ``red sequence", which have large (possibly classical) bulges (maybe formed during merger processes),  and disk galaxies in the ``blue cloud" with either no bulge or a pseudo--bulge. The red sequence population show little change in their bar fraction with luminosity and colour and overall have a high fraction of (strong) bars, approaching $50\%$. The blue cloud population also show little trend with colour with low bar fractions of 10-20\%.
  
\section{Summary}

We present here an analysis of the bar fraction of $13665$ disk galaxies selected from the new Galaxy Zoo 2 dataset. This sample is volume--limited, with $z<0.06$ and $M_r< -19.38$, and overall we find that $29.4\pm0.4\%$ of these galaxies have a bar. We split this sample as a function of global colour and luminosity, as well as the prominence of the bulge, and find that redder disk galaxies, with larger  bulges have a high fraction of bars (up to 50\%). At a fixed colour, bar fraction is seen to decrease slightly with luminosity. These results are consistent with previous visual studies of spiral galaxies (from the RC3 catalogue) as well as several recent studies \citep[][]{G10,NA10}.

We discuss the implication of our results for different scenarios of disk and bulge formation. Our results suggest a strong link between the presence of a bar and the existence of a large (possibly classical) bulge. We see hints that pseudo-bulges, if most likely found in blue, disk galaxies, will be preferentially in galaxies with no strong bar. This may be contrary to expectations which suggest pseudo--bulges are built via the re-distribution of stellar mass (or induced star-formation) driven by a bar instability while classic bulges are formed in merger like events. Furthermore, we observe a colour bimodiality in our GZ2 disk galaxies with a ``red sequence"  hosting large (possibly classical) bulges and possessing a bar fraction up to 50\%, while the majority of disk galaxies are in the ``blue cloud" which have either no bulge (or a pseudo--bulge) and possess low bar fractions of 10-20\%. These results will now need to be explained in any successful model of disk galaxy formation. 
 
This paper provides the first results from the GZ2 project on bars in disk galaxies. In the future, we will explore the dependence of bar fraction on stellar mass and environment (Skibba et al. in prep.). We will also report on a satellite Galaxy Zoo project, which invited the GZ2 users to measure the length, strength and orientation of bars (detected in the GZ2 sample) via an interactive Google Maps interface, as well as identify the links between the bar and spiral structure (Hoyle et al. in prep.). Such data will allow us to extend this work to studies of the correlation of the bar lengths (and bar colours) with global galaxy properties.

\paragraph*{ACKNOWLEDGEMENTS.} 

This publication has been made possible by the participation of more than 200,000 volunteers in the Galaxy Zoo project. Their contributions are individually acknowledged at \texttt{http://www.galaxyzoo.org/Volunteers.aspx}. KLM acknowledges funding from the Peter and Patricia Gruber Foundation as the 2008 Peter and Patricia Gruber Foundation International Astronomical Union Fellow, from a 2010 Leverhulme Trust Early Career Fellowship and from the University of Portsmouth and SEPnet (www.sepnet.ac.uk). BH thanks Google for funding during this project and RCN acknowledge financial support from STFC. Galaxy Zoo 2 was developed with the help of a grant from The Leverhulme Trust. CJL acknowledges support from an STFC Science in Society fellowship. Support for the work of KS was provided by NASA through Einstein Postdoctoral
Fellowship grant number PF9-00069 issued by the Chandra X-ray Observatory
Center, which is operated by the Smithsonian Astrophysical Observatory for and
on behalf of NASA under contract NAS8-03060. Funding for the SDSS and SDSS-II has been provided by the Alfred P. Sloan Foundation, the Participating Institutions, the National Science Foundation, the U.S. Department of Energy, the National Aeronautics and Space Administration, the Japanese Monbukagakusho, the Max Planck Society, and the Higher Education Funding Council for England. The SDSS Web Site is http://www.sdss.org/. 

KLM would like to thank Graham Smith, Irina Marinova, Sharda Jorgee and Johan Knapen for providing useful comments on the first version of this paper. KLM would also like to acknowledge the S4G bar discussion group, and particularly Eija Laurikainen for compiling an extremely useful list of references to the local bar fraction.

\end{document}